\documentclass[a4paper]{article}
\usepackage{jcappub}

\usepackage{graphicx}
\usepackage{amssymb}

\newcommand{\gagamma}{g_{a\gamma}}

\begin{document}

\title{Direct detection of dark matter axions with directional sensitivity}

\author{Igor~G~Irastorza, Juan~A~Garc\'ia}

\affiliation{Laboratorio de F\'isica Nuclear y Astropart\'iculas, Departamento de F\'isica Te\'orica, Universidad de Zaragoza, Zaragoza, Spain}

\emailAdd{Igor.Irastorza@cern.ch}

\abstract{
We study the directional effect of the expected axion dark matter signal in a resonant cavity of an axion haloscope detector, for cavity geometries not satisfying the condition that the axion de Broglie wavelength $\lambda_a$ is sufficiently larger than the cavity dimensions $L$ for a fully coherent conversion, i.e. $\lambda_a \gtrsim 2\pi L$. We focus on long thin cavities immersed in dipole magnets and find, for appropriately chosen cavity lengths, an O(1) modulation of the signal with the cavity orientation with respect the momentum distribution of the relic axion background predicted by the isothermal sphere model for the galactic dark matter halo. This effect can be exploited to design directional axion dark matter detectors, providing an unmistakable signature of the extraterrestrial origin of a possible positive detection. Moreover, the precise shape of the modulation may give information of the galactic halo distribution and, for specific halo models, give extra sensitivity for higher axion masses.
}

\maketitle

\section{Introduction}

Axions are light pseudoscalar particles that arise in theories in
which the Peccei-Quinn U(1) symmetry solves
the strong CP problem \cite{Peccei:1977hh}. They are produced in the early universe as coherent field
oscillations, the so-called misalignment mechanism~\cite{Sikivie:2006ni,Wantz:2009it}. If the PQ symmetry is
restored by reheating after inflation, axion strings and domain walls
form and decay, providing an additional source of
non relativistic axions. By these means, axions could provide all or part of the cold dark
matter of the Universe. Together with WIMPs, axions are the most attractive solution to the dark matter problem.

For these relic axions to account for the right amount of cold dark matter needed by current cosmological models, the axion mass need to be in the range $10^{-6} - 10^{-3}$ eV, the ``classic axion window''~\cite{Wantz:2009it}. Much lower $\sim$neV masses are still possible in fine-tuned models, the so-called ``anthropic axion window''
\cite{Linde:1987bx,Hertzberg:2008wr}. QCD axions with masses above the classic window are still possible, but not as a solution to the dark matter problem (unless non-standard cosmological scenarios are invoked~\cite{Visinelli2010}). Masses above 20 meV or so become in tension with current supernova modeling. Masses above $\sim$1 eV are ruled out by cosmology (too high thermal production of axions~\cite{Hannestad:2010yi}) and by astrophysical observations~\cite{Raffelt:1999tx}. Masses of a few meV could also explain the apparent anomalous energy loss of white dwarfs~\cite{Isern:2010wz,Isern1992,Isern:2008nt,Isern:2008fs}.

Axion-like particles (ALPs) or more generic weakly interacting sub-eV particles (WISPs), appearing in a number of extensions of the standard model (SM)~\cite{Jaeckel2010}, like for example string theory~\cite{Cicoli:2012sz}, may share some of the properties of the axions. Most notably, under some circumstances, they could also be a dark matter candidate~\cite{Arias:2012mb}. Some of the detection mechanisms devised for axions are applicable to other ALPs.


Most axion detection strategies invoke the a-$\gamma$-$\gamma$ interaction, present in every axion model, and that gives rise to axion-photon oscillation inside magnetic fields. Axion helioscopes~\cite{Sikivie:1983ip} look for the large axion flux emitted by our Sun with keV energies, using a powerful magnet pointing to the Sun. The most powerful helioscope built, CAST~\cite{Zioutas:2004hi,Andriamonje:2007ew,Arik:2008mq,Aune:2011rx}, is currently sensitive to QCD axion models at the $\sim$ eV scale. The future IAXO~\cite{Irastorza:2011gs} will have a sensitivity to scan a large fraction of the models in the high mass range $10^{-3}-1$ eV. ALPs can also be searched for purely in the laboratory~\cite{Ehret:2010mh}, but with insufficient sensitivity to reach axion models.

In the classic window, and under the assumption that axions are the dominant component of dark matter, relic axions can be directly detected using Sikivie's haloscope technique~\cite{Sikivie:1983ip}. Due to the relic axions being non-relativistic, axion conversion gives photons with energies equal to the axion mass, in the microwave regime. If the conversion happens in a microwave cavity resonant to the axion mass, due to the low velocity dispersion of the axions, the conversion is substantially enhanced, and a sufficiently high sensitivity can be obtained to explore realistic QCD axion models. This technique has been used already by a number of experiments, being ADMX the most powerful haloscope to-date \cite{Asztalos:2001tf,Asztalos:2003px}, with sensitivity to QCD axions of $\mu$eV mass.

According to the standard formalism of the haloscope technique~\cite{Sikivie:1983ip}, the power delivered into a cavity immersed in a magnetic field due to conversion from relic axions is:

\begin{equation}
    P_0 = \gagamma^2 V B^2  C \frac{\rho_a}{m_a} Q
\label{pout}
\end{equation}

\noindent where $V$ is the volume of the cavity, $B$ its magnetic field strength, $Q$ its the loaded quality factor of the cavity (that we have assumed that it is lower than the relative energy spread of the axion energy $Q_a \sim 10^6$), and $C$ is a geometry factor involving the precise electric field of relevant resonant mode in the cavity $\textbf{E}_{\textrm{cav}}(\textbf{x})$ and the magnetic field $\textbf{B}(\textbf{x})$:

\begin{equation}
   C= \frac{\left(\int dV \textbf{E}_{\textrm{cav}}(\textbf{x}) \textbf{B}(\textbf{x})\right)^2}{V|\textbf{B}|^2 \int dV \epsilon(\textbf{x}) \textbf{E}^2_{\textrm{cav}}(\textbf{x})}
\label{cfactor}
\end{equation}

The previous equations are valid under the basic assumption that the typical de Broglie wavelength of the relic axions $\lambda_a$ is longer that the characteristic size of the cavity $d$.

\begin{equation}\label{condition}
   \lambda_a \gtrsim d
\end{equation}

Only in this case the resonant conversion giving rise to (\ref{pout}) takes place. The axions composing the dark matter halo have a velocity distribution given by the specific halo model but in general with typical values of $\sim$ 300 km/s. So approximately the de Broglie wavelength of the relic axions depends on the axion mass in the following way:

\begin{equation}\label{debroglie}
 \lambda_a = \frac{2\pi}{p_a} \sim 12.4 \textrm{ m} \left(\frac{10^{-4}\textrm{ eV}}{m_a}\right) \left(\frac{300 \textrm{ km/s}}{v_a}\right)
\end{equation}

As can be seen from (\ref{debroglie}), for axion masses well below 10$^{-4}$ eV, any magnet geometry in the $\sim$m scale is well within the condition (\ref{condition}). ADMX is indeed using a cylindrical cavity of 1 m length and 0.6 m diameter inside a solenoidal magnet, tunable to axion masses at the few $\mu$eV scale, and enjoys sensitivity sufficient to exclude one of the two main axion benchmark models (KSVZ axions)~\cite{Kim:1979if, Shifman:1979if}. Improvements to enhance the sensitivity to lower $\gagamma$ values are ongoing as well as R\&D to build setups resonant at higher axion masses (above $10^{-5}$ eV)~\cite{Asztalos2010,2012APS..APRD14001M}. For higher axion masses, the main challenge relies on the fact that the needed resonant cavity geometries are smaller and so is their corresponding sensitivity ($P$ being proportional to the volume of the cavity $V$). Recently the use of long thin cavities (waveguides) inside strong dipole magnets have been proposed as a possibility to achieve competitive sensitivity in the 10$^{-5}$-10$^{-4}$ eV range~\cite{Baker:2011na,Caspers:2010zz}. The use of small cross-section, but long and powerful, dipole magnets like the ones used in particle accelerators (and already recycled for axion physics in, e.g., CAST~\cite{Aune:2011rx} or ALPS~\cite{Ehret:2010mh}) can accommodate cavities resonant at these higher frequencies (driven by the small dimension of the waveguide) while at the same time keeping large enough $V$ and $B$. Moreover, this option may give additional technical advantages with respect the conventional approach, e.g., regarding mode crossing or mode localization~\cite{Baker:2011na}.

The possibility of using few-meter long cavities for detection of few $\times 10^{-5}$ eV relic axions bring us closer to the limitation expressed by~(\ref{condition}). In this paper we explore in some detail the effect of this limitation on the predicted relic axion signal for realistic dark matter momentum distributions. In particular, we note that for thin and long geometries like the ones mentioned before, the orientation of the cavity with respect the main incoming axion direction affects the signal intensity. For carefully chosen lengths this effect can be maximized and used as a powerful identification signature of the extraterrestrial origin of an eventual positive detection. Moreover, like in the case of WIMP directional experiments~\cite{Ahlen:2009ev},
the resulting signal modulation will show specific features of the galactic dark matter halo, enabling us to discriminate between different options currently considered. 

Finally, we want to stress that directionality in ALP detection (and emission) antennas were first discussed in~\cite{Caspers:2010zz}, where concepts like phased arrays, and excitation of higher order or traveling wave modes were put forward. They were qualitatively discussed in the context of microwave-shinning-through-wall experiments sensitive to hidden photons and ALPs, for which directionality would give extra sensitivity, although of course also usable in relic axion searches. Here we focus ourselves on the relic axion case, and on the simple setup consisting on a long rectangular cavity in a constant magnetic field and resonant in a fundamental mode. As mentioned, already in such simple case directional effects appear when the condition~(\ref{condition}) is not satisfied. We will compute the signal strength in that particular situation, study the directional effect and derive prescriptions for the length of the cavity needed to maximally exploit it. We will convolute the result with the momentum distribution of relic axions at the Earth, according to two different halo models, in order to assess its potential in realistic conditions and perform some preliminary sensitivity computation. Of course these concepts could be generalized to arbitrary cavity geometries, that could evolve into true relic axion antennas, however, the restricted scenario studied here is specially appealing because of the feasibility of constructing such setups in the near future.

The article is structured as follows. In section \ref{sec:signal} we discuss the basic concept for an idealized situation with a single axion incoming direction. In section \ref{sec:halo} we present our numerical calculations for a realistic axion momentum distribution following the standard isothermal isotropic model. In section \ref{sec:caustics} we present the calculations for a particular halo model, that of Sikivie's caustic rings. In section~\ref{sec:sensitivity} we perform some sensitivity calculations. We finish with our conclusions in section \ref{sec:conclusions}.

\section{Directional effect}
\label{sec:signal}

As noted before, expression (\ref{pout}) is calculated assuming that de Broglie wavelength of the incoming axion is larger than the dimensions of the cavity. Thus the axion is approximated by a spatially constant oscillating field. If we relax this condition, the spatial variation of the axion field along the cavity volume cannot be neglected. In general we can express the new result as:

\begin{figure}[b]
\centering
\includegraphics[width=6cm]{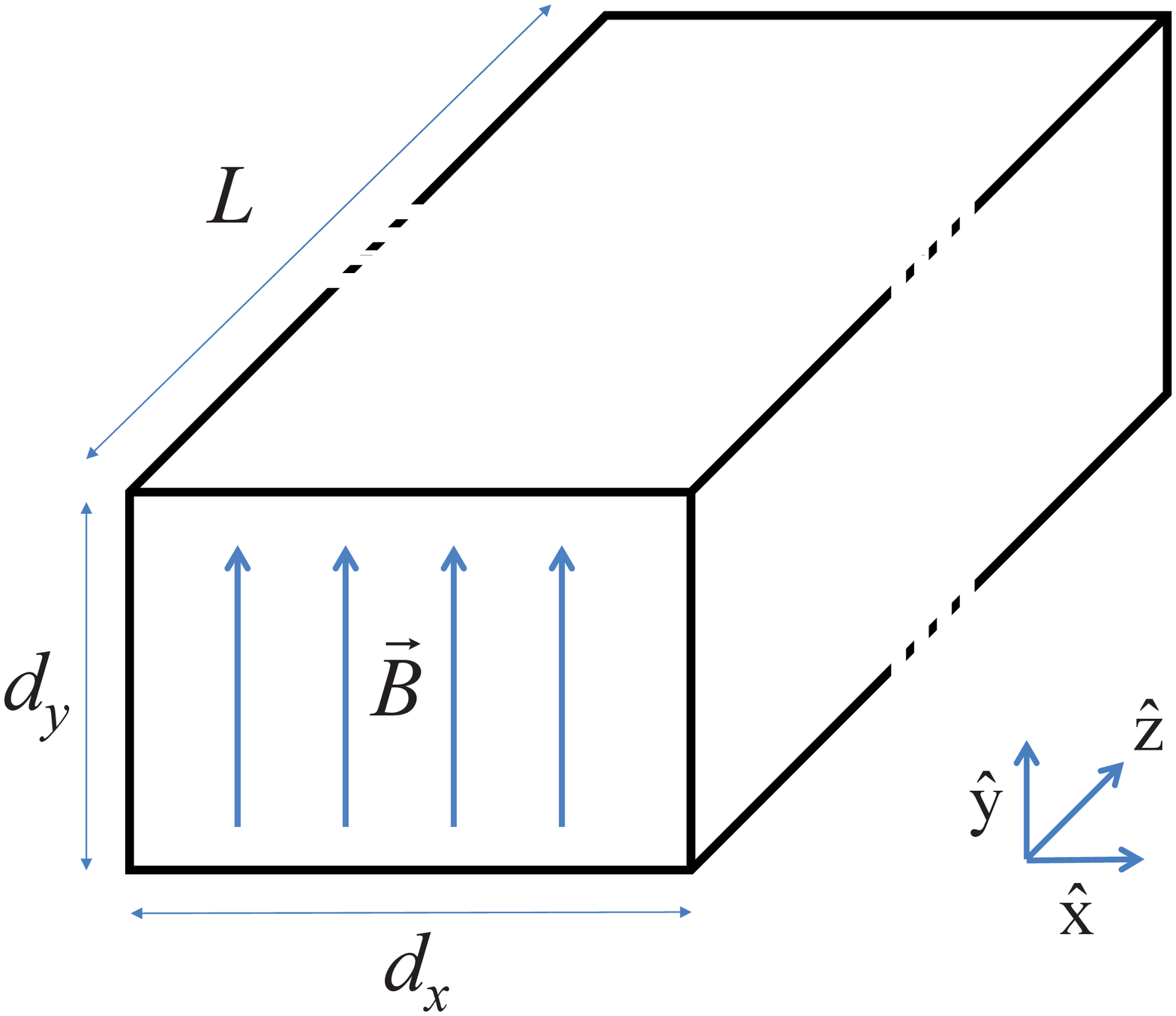}
\includegraphics[width=9cm]{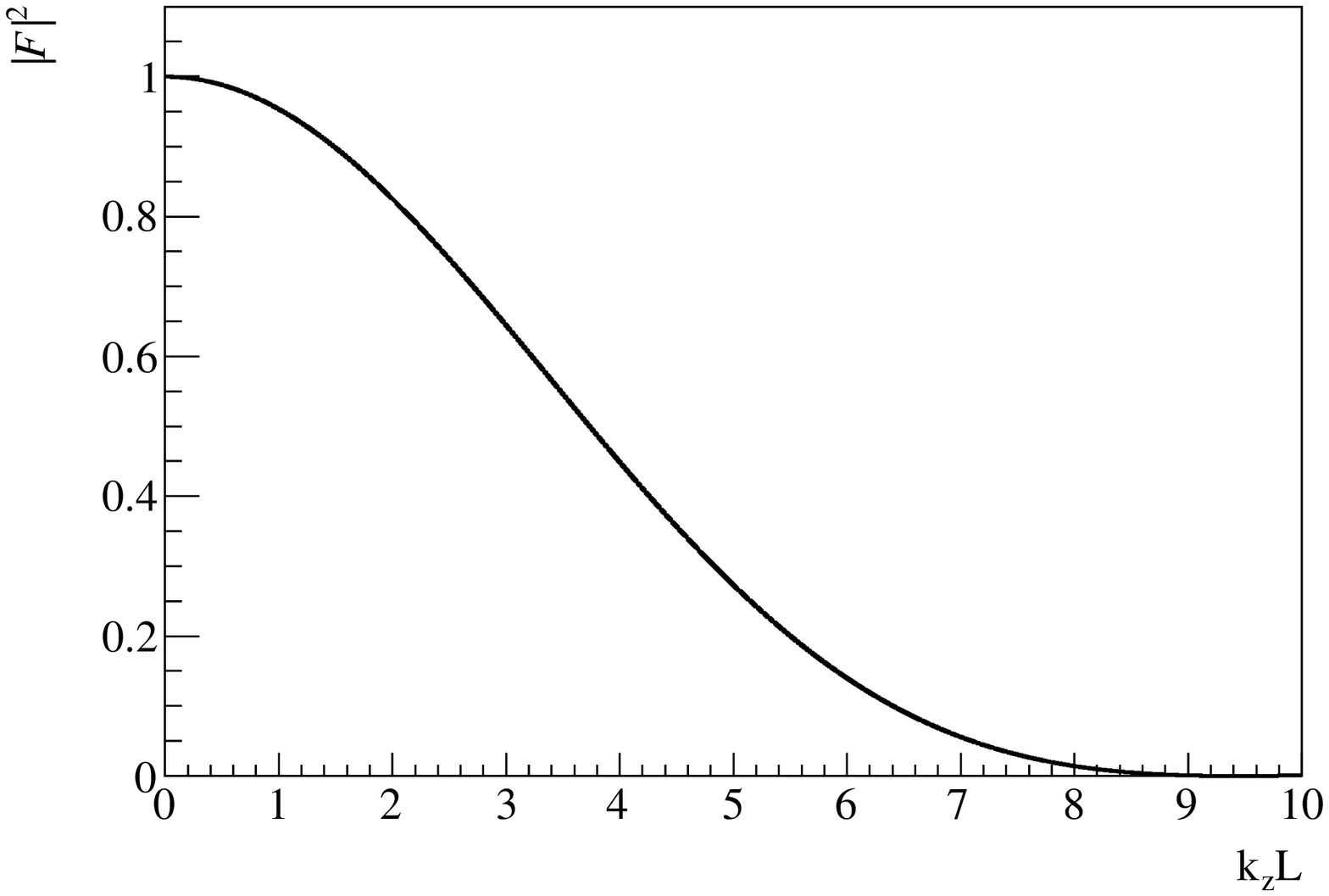}
\caption{Left: sketch of the cavity geometry used for the calculations. Right: $|\mathcal{F}|^2$ versus $k_zL$ for a single incoming axion momentum $k$ as expressed in eq. (\ref{ctilde}).}
\label{fig:fk}
\end{figure}

\begin{equation}\label{pout2}
    P_{out} = \gagamma^2 V B^2  \tilde{C} \frac{\rho_a}{m_a} Q
\end{equation}

\noindent where the new constant $\tilde{C}$ (in general $\tilde{C} \leq C$) includes now a more complex dependency on the axion incoming momenta and the cavity geometry and field.

We focus on the specific case of a thin and long (in the $\mathbf{\hat{z}}$ direction) rectangular resonant cavity, inside a dipole magnet of constant $B$ field parallel to $\mathbf{\hat{y}}$, i.e., along one of the small dimensions of the cavity.  The only relevant cavity mode is then the TE$_{101}$ mode, the most fundamental one with the $E$ field parallel to $B$. This approximation assumes that the typical axion wavelengths may be comparable or shorter than the length $L$ of the magnet but otherwise the condition~(\ref{condition}) is preserved, however, for the short dimensions of the cavity $d_x$, $d_y$. In this approximation and for the idealized case of a single incoming axion direction and momentum $k$, the expression for $\tilde{C}$ can easily be computed analytically. For convenience, we express $\tilde{C} = C |\mathcal{F}|^2$, where $\mathcal{F}$ is a form factor expressing the loss of coherence due to the axion momentum along the length of the cavity:

\begin{equation}\label{fk}
    \mathcal{F}(k) = \frac{\pi}{2L}\int_L e^{ik_z z} \sin \left(\frac{\pi z}{ L}\right) dz
\end{equation}

\noindent where $k_z$ is the projection of the axion momentum along the length of the cavity. The expression for $\tilde{C}$ yields:

\begin{equation}\label{ctilde}
    \tilde{C} = C |\mathcal{F}(k)|^2 = C \frac{(1+\cos(k_z L))\pi^4}{2(\pi^2-k_z^2L^2)^2}
\end{equation}

This expression is plotted in figure \ref{fig:fk} where the anticipated behavior is clearly seen. It provides the conventional result $\tilde{C} = C$ for low values of $k_zL \lesssim 1$ while the signal drops in intensity for larger values of $k_zL$. We must note that low values of $k_zL$ are achieved by small $L$ or $k$, but also for axion incoming directions perpendicular to the cavity length. This suggests that full coherence is possible even for cavities longer that de Broglie wavelength, provided the approximation of long thin cavity is valid and we are able to orient the cavity perpendicular to the main axion direction. This could give an extra sensitivity bonus to long thin cavities by allowing to use longer cavities that partially compensate the relatively lower volumes. However, a more realistic treatment using a distribution of axion incoming directions needs to be done to study the real extent of this assertion. This is addressed in the next two sections.

\section{Isothermal sphere halo model}
\label{sec:halo}

\begin{figure}[b]
\centering
\includegraphics[width=7.5cm]{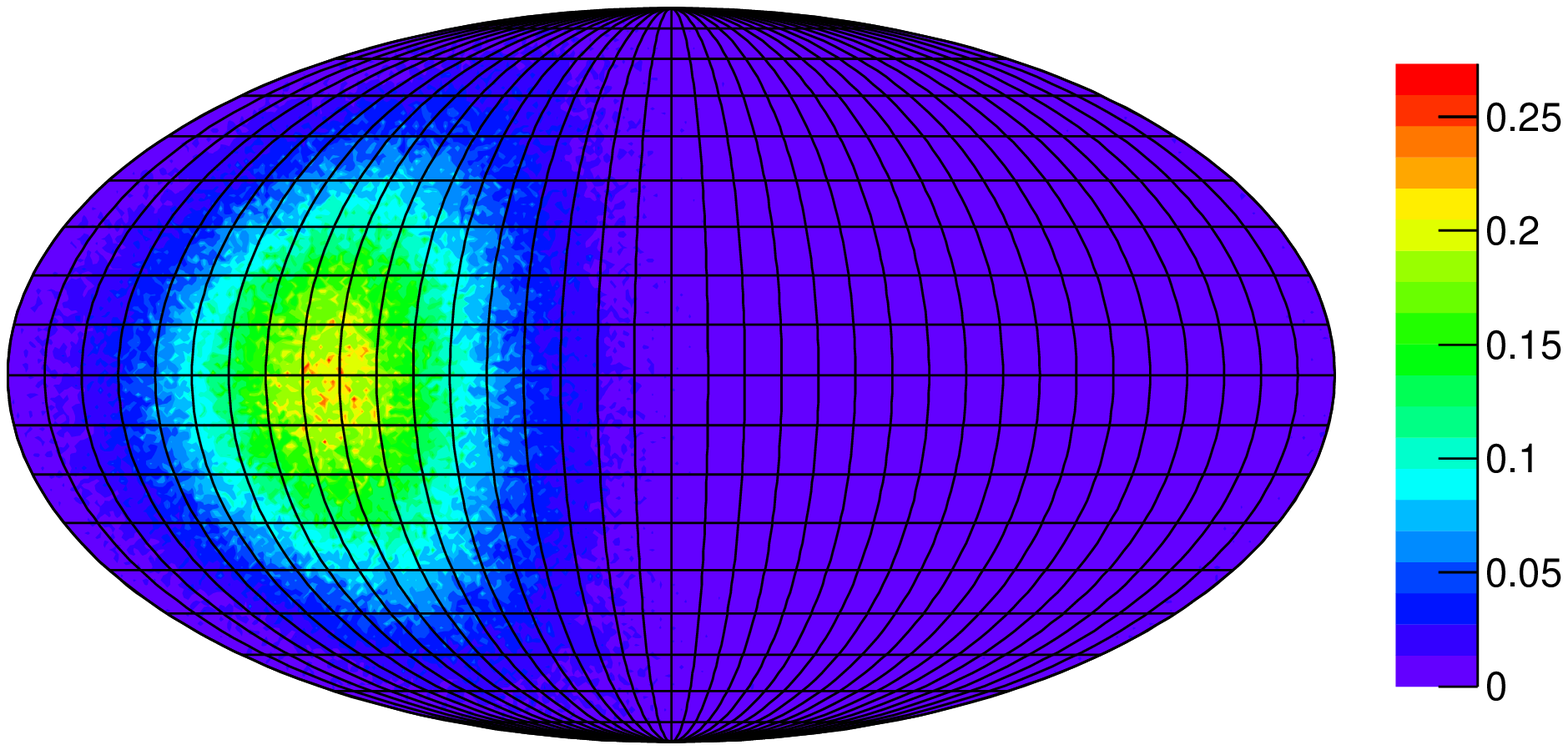}
\includegraphics[width=7.5cm]{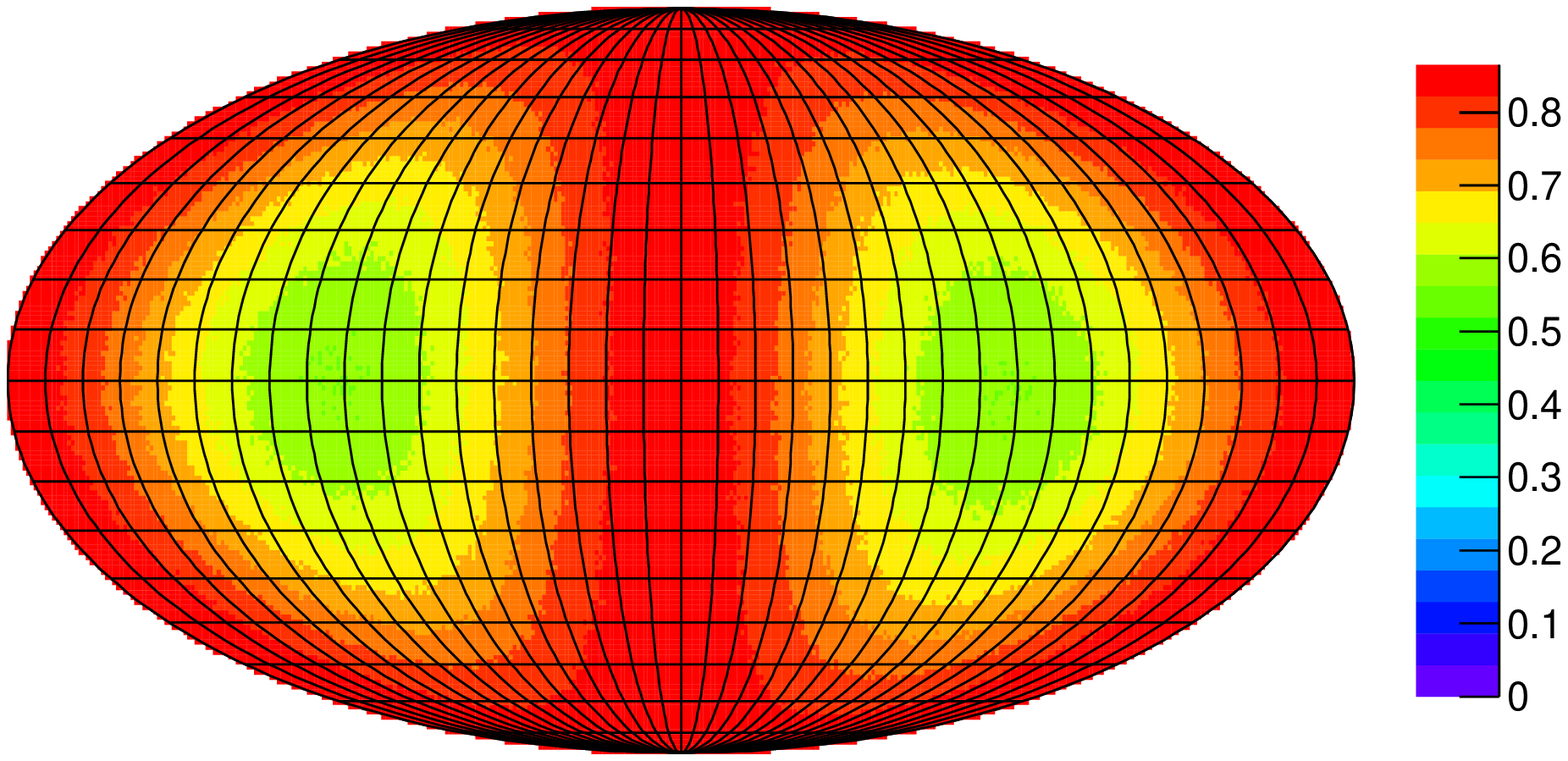}
\caption{(Left) Mollweide projection of the velocity distribution predicted by the isothermal sphere halo model. (Right) Value of  $|\mathcal{F}_{\mathrm{IS}}|^2$ (i.e. proportional to the axion signal) for all possible 2-angle orientation of the cavity using the Mollweide projection in galactic coordinates, for the particular case of $m_a=6 \times 10^{-5}$ eV and $L=15$ m. The signal is minimum when the cavity is oriented to the CYGNUS point and maximum when it is perpendicular to it.}
\label{fig:isothermalsphere}
\end{figure}

The velocity distribution of dark matter axions at the Earth depends on the assumptions and details considered for the halo model, i.e., the way the dark matter is distributed in the galaxy. The halo model must be consistent with the observed rotation curve of the galaxy. Nevertheless, while keeping this main constraint, a large number of different halo models can be considered, and there is a vast literature on the issue. More specifically the velocity distributions of the dark matter particles at Earth predicted by each halo model has been determined and studied in the context of WIMP dark matter experiments. We refer, e.g. to~\cite{PhysRevD.66.043503} for a description of possible halo models and their corresponding velocity distributions.

The simplest one of these halo models is the isothermal sphere model, where the dark matter particles follow a spherical distribution with a flat rotation curve. The distribution in velocity space turns out to be isotropic and Maxwellian:

\begin{equation}\label{DF}
 f(\vec{v}) = f(v) \propto \exp (-3v^2 / (2v^2_{\textrm{rms}}))
\end{equation}

\noindent where $ v = |\vec{v}| $ and $v_{\textrm{rms}}$ is the root mean square velocity $v_{\textrm{rms}} = \sqrt{3/2}v_0$ being $v_0$ the rotation speed of the galaxy at the solar system radius. This is used as a benchmark standard model in the calculation of expected rates and derivation of exclusion plots in direct WIMP dark matter detection experiments, although its choice is due more to its simplicity rather than its physics motivation. The underlying assumptions of sphericity, non-rotation of the dark matter halo, isotropy, etc, are not totally sustained on physics grounds or observations. Therefore, many families of alternative galactic halo models have been considered, each one introducing a deviation from the standard isothermal sphere. Models with non-isotropic velocity dispersions, with non-spherical distributions (axisymmetric or triaxial), or with some degree of corotation or counterrotation, have been considered in the literature~\cite{PhysRevD.66.043503} in the context of studies of systematic uncertainties in WIMP direct detection rates.

In addition to the particular shape of $f(v)$ due to the particular halo model considered, one has always to consider the effect of the movement of the Earth-Sun system through the galactic dark matter halo. Expressed in the terrestrial frame of reference, the velocity distribution function $f(\vec{w})$, being $w$ the incoming velocity of the axions at Earth, is derived from $f(\vec{v})$ after relating $\vec{v} = \vec{w} + \vec{u}$ where $\vec{u}$ is the velocity of the Earth in the galactic rest frame. This translation produces a general anisotropy in the velocity distribution at the Earth frame of reference, centered in the point in the sky (the CYGNUS point\footnote{In galactic coordinates, approximately $b=0^\circ, l=90^\circ$.}) to which $\vec{u}$ is directed to. Let us comment too that $\vec{u}$ is of the order 220 km/s and has a oscillatory component of about 12 km/s due to the rotation of the Earth around the Sun.

The distribution $f(\vec{w})$ corresponding to the isothermal sphere model of eq.~(\ref{DF}) is shown on the left of figure \ref{fig:isothermalsphere} in the typical equal-area Mollweide projection and galactic coordinates. As can be seen, it is dominated by the mentioned anisotropy introduced by the Earth-Sun motion. Although for other halo models the particular distribution will differ, this general anisotropy is expected to be present for every model unless a substantial corotation of the dark matter halo is assumed, something that does not seem very favored due to the low interaction between dark and conventional matter.

\begin{figure}[t]
\centering
\includegraphics[width=7.5cm]{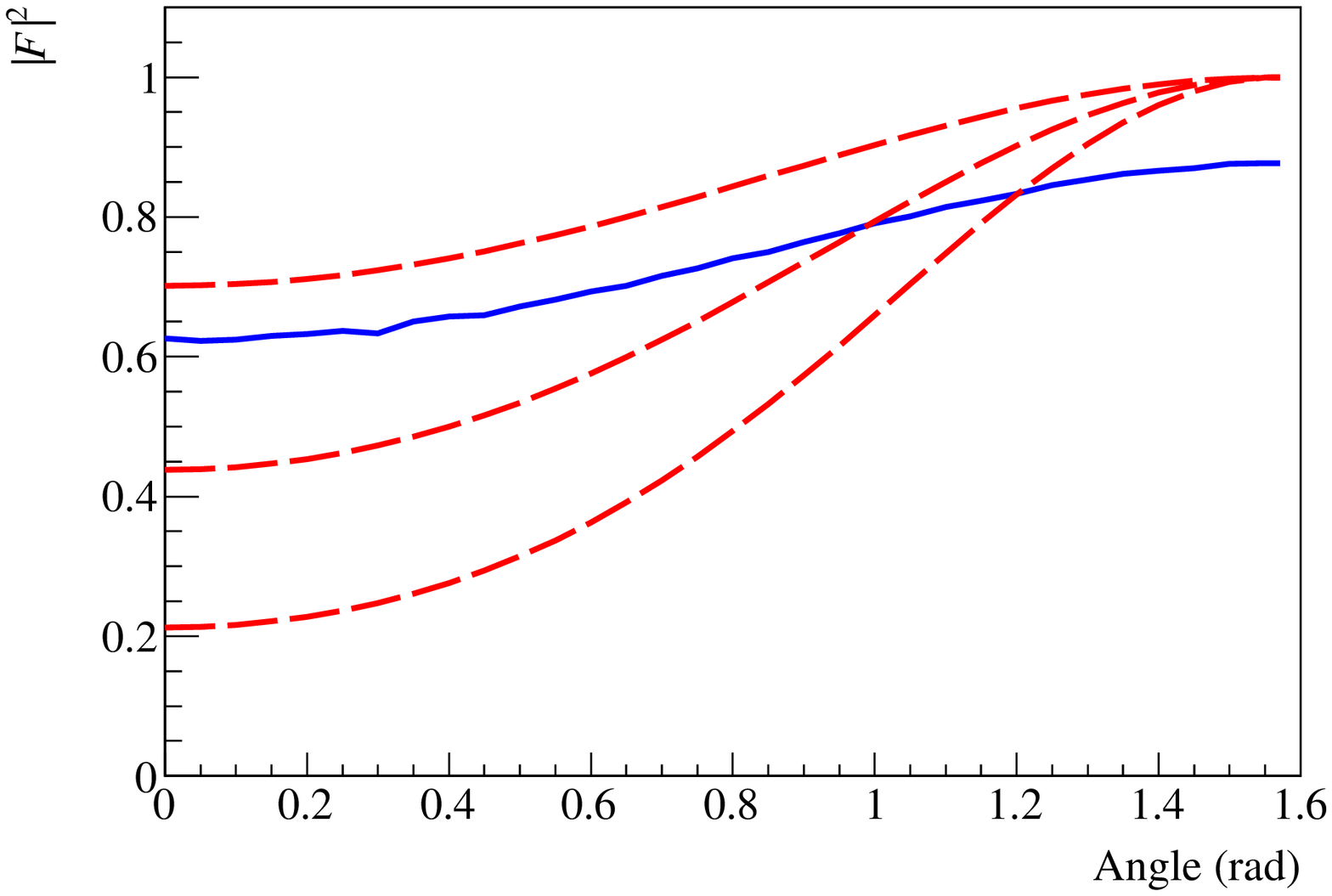}
\includegraphics[width=7.5cm]{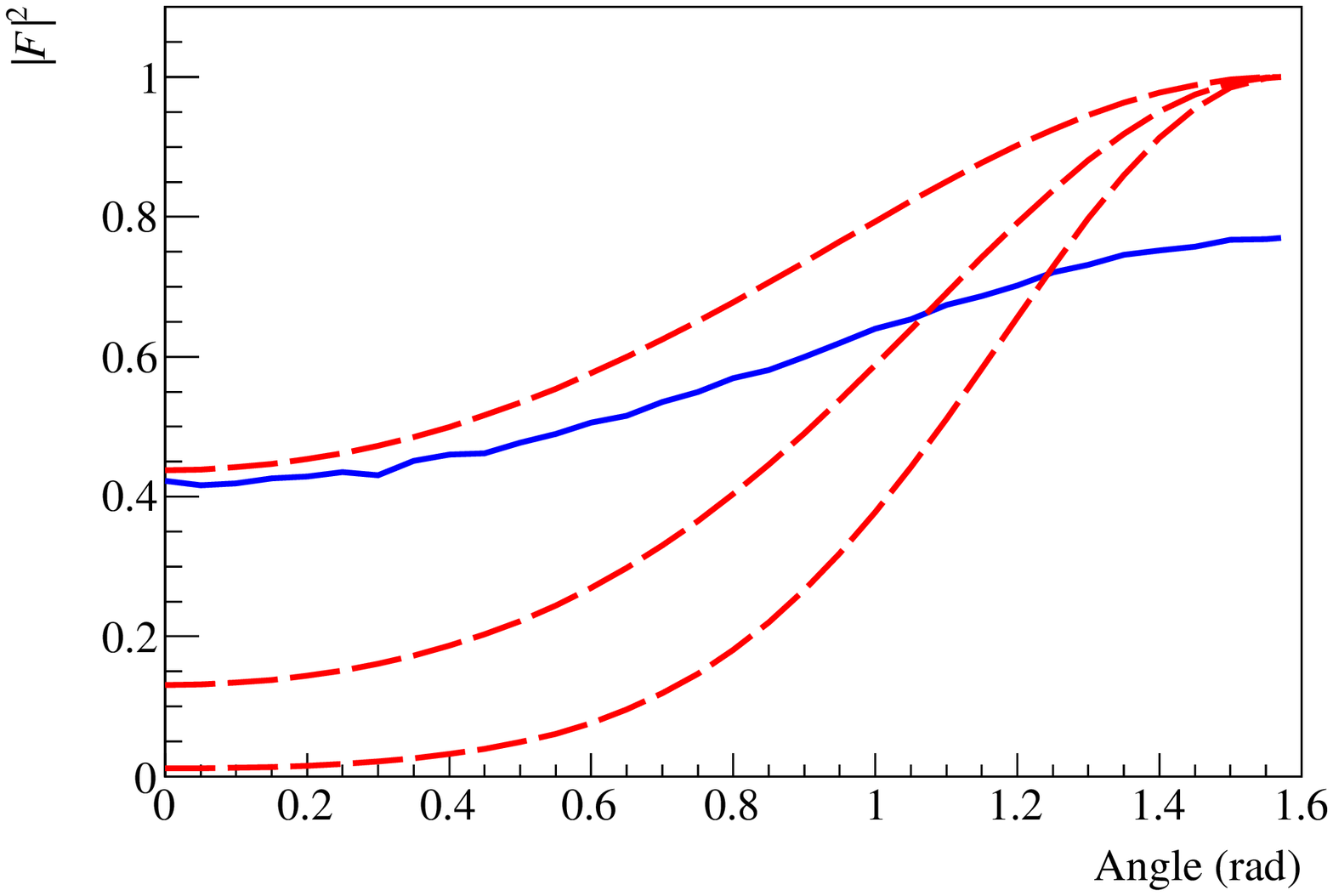}
\caption{Factor $|\mathcal{F}_{\mathrm{IS}}|^2$ versus magnet orientation with respect the CYGNUS point (solid blue line) compared with the idealized situation of a single-direction incoming axion according to equation~\ref{ctilde}, (red dashed lines, corresponding, from top to bottom, to $v_a =$ 200, 300 and 400 km/s). The left plot corresponds to $m_{a} = 4 \times 10^{-5}$~eV  and $L = 20$~m and  the right one to $m_{a} = 8 \times 10^{-5}$~eV and $L = 15$~m.}
\label{fig:fkis}
\end{figure}

Using this benchmark distribution, we have calculated numerically a more realistic version of the constant $\tilde{C}$ by convoluting
the form factor introduced in~(\ref{fk}) with the momentum distribution provided by the isothermal sphere model~(\ref{DF}):


\begin{equation}\label{FIS}
  | \mathcal{F}_{\mathrm{IS}} |^2 = \int_k f(k) | \mathcal{F}(k)|^2 dk
\end{equation}

The result is illustrated in figure~\ref{fig:fkis}, where the factor $|\mathcal{F}_{\mathrm{IS}}|^2$ is plotted versus the cavity orientation angle, for two choices of axion mass $m_a$ and cavity length $L$. The orientation angle is defined with respect the CYGNUS point in the sky, i.e. the center of the axion velocity distributions. The curves are compared with idealized curves assuming a monochromatic single-direction axion beam using expression~(\ref{ctilde}). As expected, the momentum dispersion of distribution (\ref{DF}) causes a smoothing of the modulation of the signal with orientation, compared with the idealized cases. However, a significant O(1) modulation still remains for some values of $m_a$ and $L$. In addition, the signal at the maximum decreases for longer cavities with respect to full coherence ($|\mathcal{F}_{\mathrm{IS}}|^2=1$), because the momenta dispersion in the distribution prevents to achieve the condition $k_zL=0$ of the idealized case. On the right of figure~\ref{fig:isothermalsphere}, $|\mathcal{F}_{\mathrm{IS}}|^2$ is plotted for all possible 2-angle orientation of the cavity for the particular case of $m_a=6 \times 10^{-5}$ eV and $L=15$ m.

\begin{figure}[t]
\centering
\includegraphics[width=11cm]{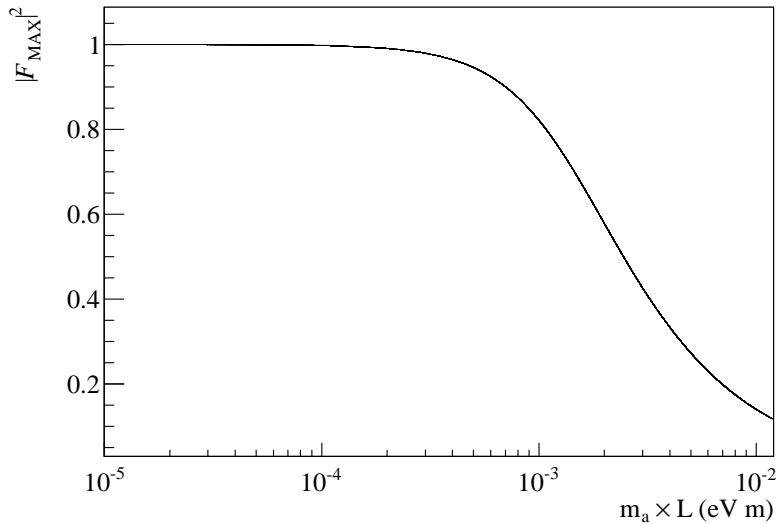}
\caption{Value of $|\mathcal{F}_{\mathrm{IS}}|^2$ versus $m_a L$ for the maximum of the modulation, i.e., with the cavity oriented perpendicular to the CYGNUS point. We observe that as long as the cavity length is not longer than $\sim5\times (10^{-4} \mathrm{eV} / m_a) $ meters, full coherence is kept and the conventional expression for the signal is valid, provided the orientation of the cavity is kept. Slightly longer magnets still give signal but in decreasing intensity, for example, for lengths of $\sim20\times (10^{-4} \mathrm{eV} / m_a) $ meters, the signal drops at $\sim$60\%.}
\label{fig:maxC}
\end{figure}

\begin{figure}[t]
\centering
\includegraphics[width=11cm]{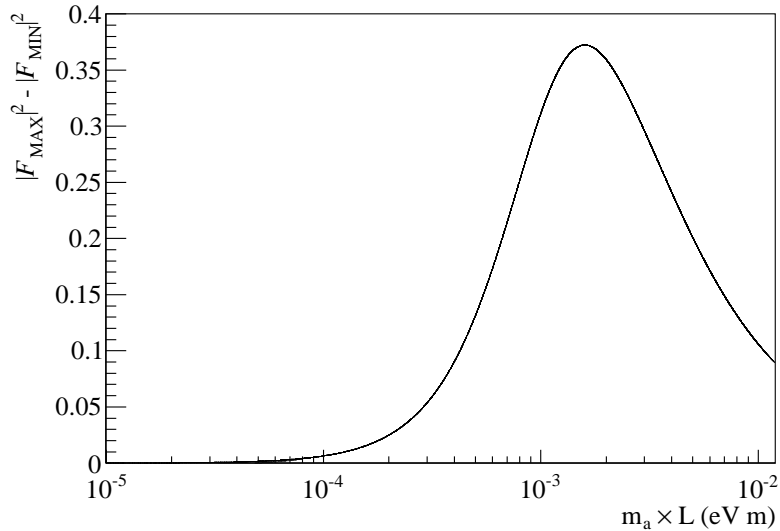}
\caption{Difference of the values maximum and minimum of $|\mathcal{F}_{\mathrm{IS}}|^2$ (i.e., with the cavity oriented perpendicular and parallel to the CYGNUS point, respectively) versus $m_a$ and $L$. Maximum difference is obtained for $m_aL \sim 10-25 \times 10^{-4}$eV m, showing the interesting area to exploit the directional signal.}
\label{fig:diffCmaxmin}
\end{figure}

Figures~\ref{fig:maxC} and~\ref{fig:diffCmaxmin} show this issue in a more systematic way. Fig.~\ref{fig:maxC} plots $|\mathcal{F}_{\mathrm{IS}}|^2$ at the maximum (i.e. cavity oriented always perpendicular to the CYGNUS point) for a range of values of $m_a L$. Full coherence is preserved up to a given (axion mass dependent) length of the cavity. For cavities shorter or approximately equal to $\sim5\times (10^{-4} \mathrm{eV} / m_a) $ meters, full coherence is kept and the conventional expression for the signal is valid (provided the orientation of the cavity is kept). Slightly longer magnets still give signal but in decreasing intensity with respect the one expected for their volume, for example, for lengths of $\sim10\times (10^{-4} \mathrm{eV} / m_a) $ meters, the signal drops at $\sim$60\% of $P_0$.

Figure~\ref{fig:diffCmaxmin} plots the difference of $|\mathcal{F}_{\mathrm{IS}}|^2$ at the maximum and the minimum of the modulation (i.e., with the cavity oriented perpendicular and parallel to the CYGNUS point, respectively). It is interesting to see that for an adequately chosen length of the cavity, a modulation difference as large as 35\% of $P_0$ can be expected. This happens for lengths of $\sim10\times (10^{-4} \mathrm{eV} / m_a) $ m, which, according to figure~\ref{fig:maxC} correspond to geometries at the transition region where full coherence is lost but still the expected signal is 60-70\% of the full coherence value $P_0$. This region should be noted as a desirable operating point, providing almost full coherent signal when the cavity is oriented perpendicular to the CYGNUS point, while showing a strong (50\% relative) modulation with the cavity orientation, providing a strong identificative signature.

\section{Low dispersion streams}
\label{sec:caustics}

Especially interesting are the halo models with some degree of clustering in velocity space. In particular, streams of dark matter in the local neighborhood could be formed by tidal disruption of dwarf satellites or due to the late infall of dark matter onto the galaxy. These streams would have a much lower velocity dispersion than a thermal component of the kind studied in the previous section. There is no general consensus on the presence, type or importance of dark matter streams in the Milky Way and in particular which fraction of the local dark matter density is in the form of streams. The issue is however or great importance in the prediction of signal rate in direct WIMP detectors. Local low dispersion WIMP streams would produce characteristic sharp rises in the nuclear recoil spectra in WIMP detectors~\cite{PhysRevD.71.043516}. In axion haloscopes, axion dark matter streams would manifest themselves as very narrow peaks in frequency. ADMX has looked for those in special high resolution runs~\cite{Duffy:2005ab}. These streams would also have very important consequences regarding the effect studied here. They could leave specific imprints in the dependence of the signal with cavity orientation. Moreover, for low dispersion components the coherence condition gets relaxed, allowing for full signal conversion at higher values of $m_a L $, provided the orientation of the cavity is perpendicular to the stream direction.

For the sake of concreteness, we focus now on the infall self-similar model developed by Sikivie and collaborators~\cite{Duffy2008,Natarajan:2005fh}. This model is extreme in the sense that it predicts that most of the local dark matter density is in the form of streams. The model describes the dark matter as a collisionless fluid that falls into the galactic gravitational well, and starts oscillating around it without losing its condition of a thin sheet in phase space. In the turnaround areas high density structures, called caustic rings, form, for which some observational evidence has been suggested~\cite{Sikivie:2001fg}. Locally the dark matter is in the form a discrete set of flows, each of them characterized by the number of times they have bounced back in the galactic well. The proponents of this model have predicted the specific direction, density and velocity of the first 40 flows at the Earth location~\cite{Duffy2008}. It turns out that two thirds of the local dark matter density is carried by one single flow, dubbed the Big Flow~\cite{Duffy2008}. The velocity dispersion of these flows can be very small. Axions crossing the gravitational structure of the galaxy will get very slightly dispersed $\delta v / v \lesssim 0.001$ ~\cite{Sikivie1992}. We therefore assume a very conservative value for the velocity dispersion of each flow $\delta v_n$, where $n$ is the flow number:

\begin{equation}\label{deltav}
 \delta v_n = n \textrm{\ km/s}
\end{equation}

\noindent although a much lower dispersion of $\sim50$ m/s has been proposed for the Big Flow on observational grounds~\cite{Sikivie:2001fg}. The more conservative value (\ref{deltav}) seems convenient for our case study because it is also closer to the values, of order $\sim$10 km/s, assumed for the velocity dispersion of dark matter streams produced by tidal disruption of dwarf satellites.

\begin{figure}[b]
\centering
\includegraphics[width=7.5cm]{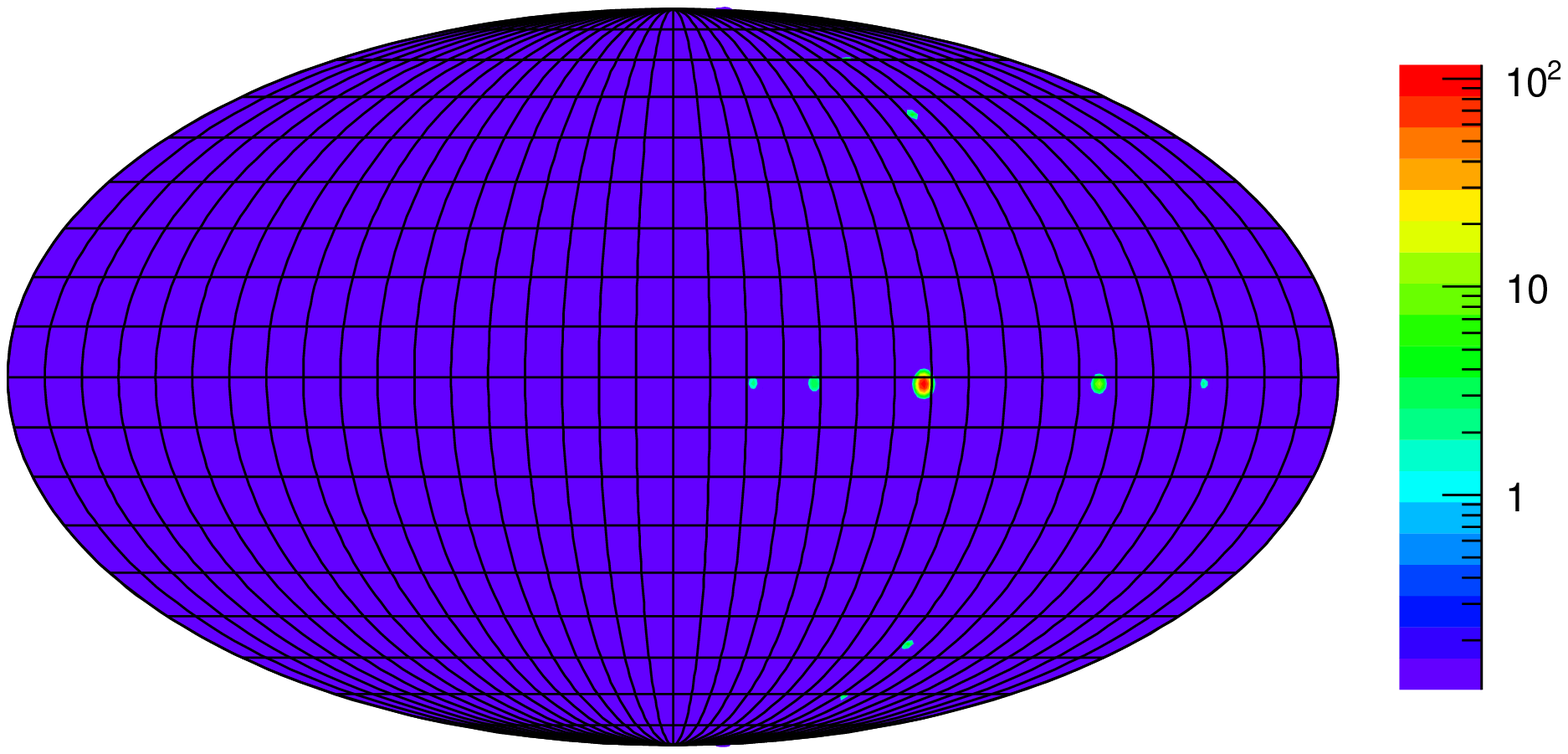}
\includegraphics[width=7.5cm]{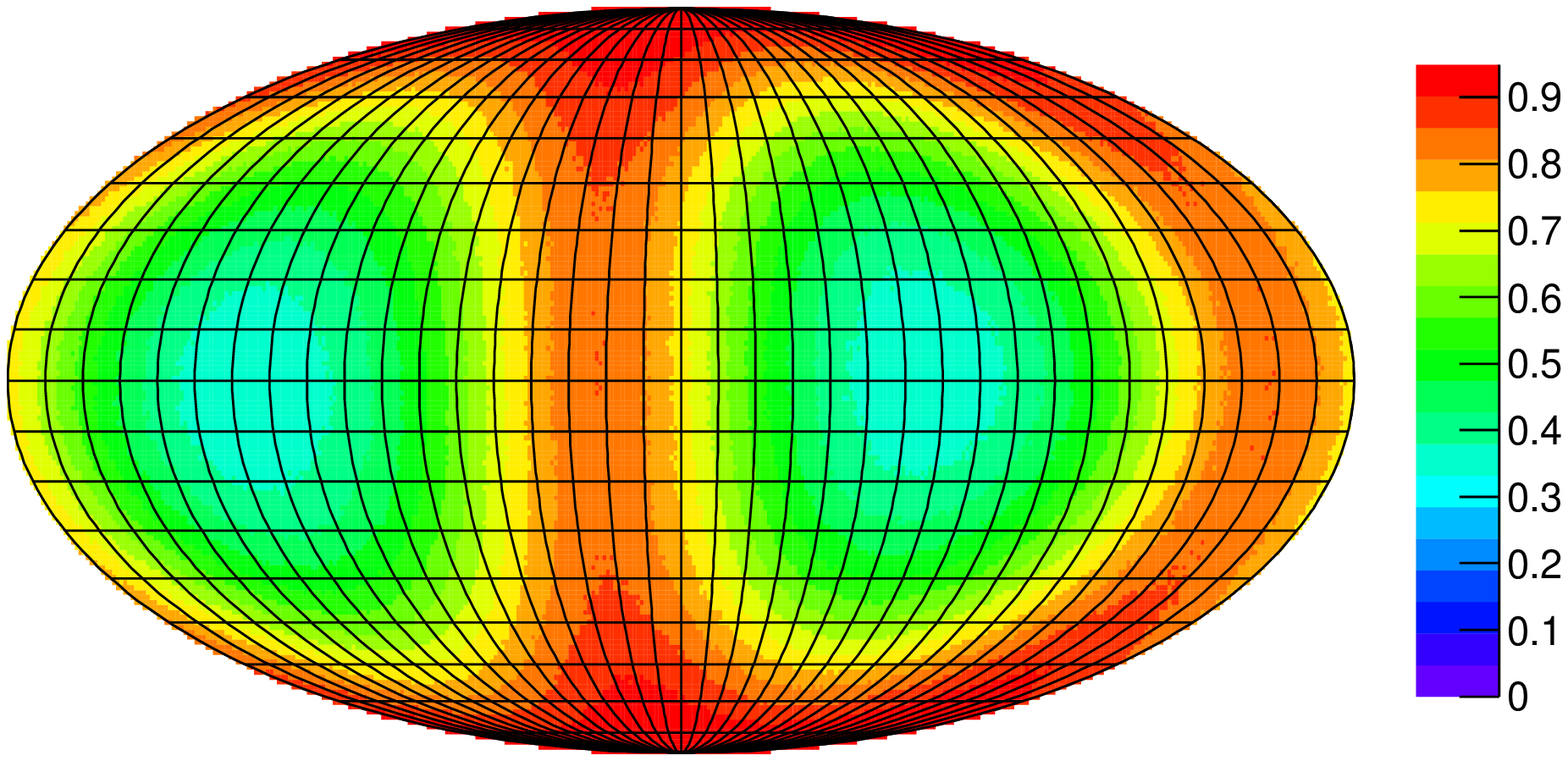}
\caption{(Left) Mollweide projection of the velocity distribution predicted by the late infall self-similar model discussed in the text, with the velocity dispersion per flow indicated by~(\ref{deltav}). (Right) Value of  $|\mathcal{F}_{\mathrm{IM}}|^2$ for all possible 2-angle orientation of the cavity using the Mollweide projection in galactic coordinates, for the particular case of $m_a=7 \times 10^{-5}$ eV and $L=15$ m. The signal is minimum when the cavity is oriented close to the Big Flow direction and maximum when it is perpendicular to it. }
\label{fig:causticmap}
\end{figure}

\begin{figure}[t]
\centering
\includegraphics[width=11cm]{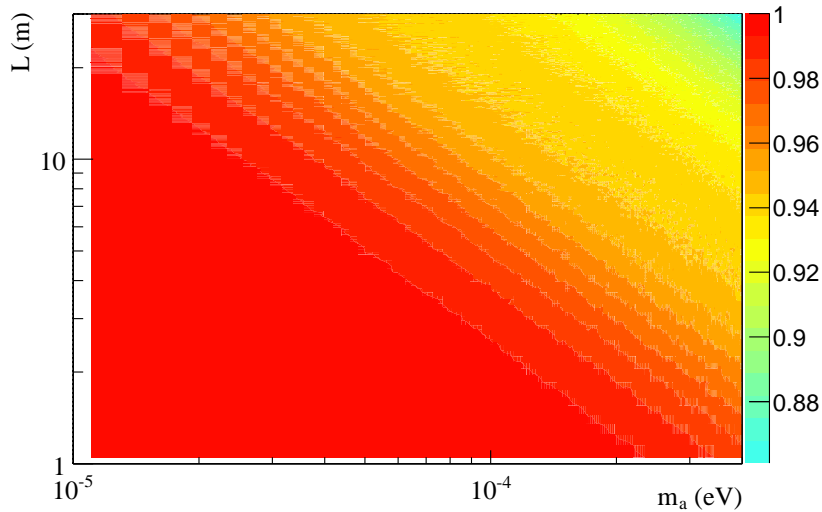}
\caption{Value of the form factor (\ref{fk}) for the caustics rings model, $|\mathcal{F}_{\mathrm{IM}}|^2$, versus $m_a$ and $L$ for the maximum of the modulation.
}
\label{fig:maxCcaustics}
\end{figure}

\begin{figure}[t]
\centering
\includegraphics[width=11cm]{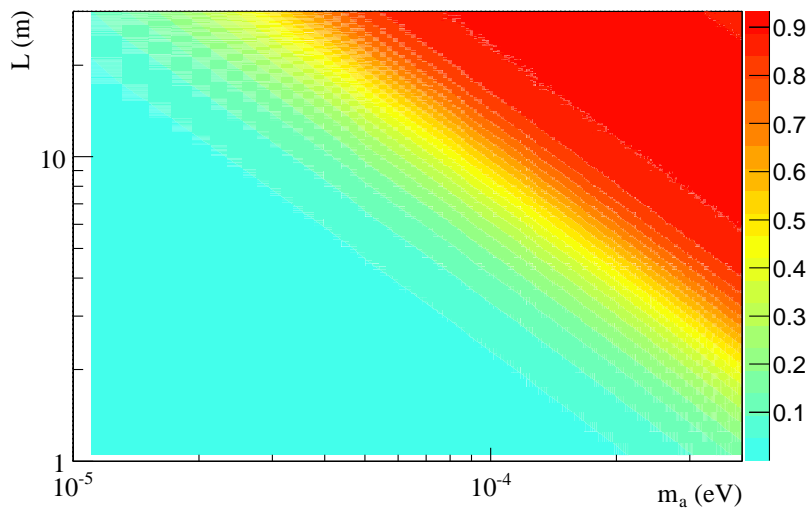}
\caption{Difference of the values maximum and minimum of $|\mathcal{F}_{\mathrm{IM}}|^2$ versus $m_a$ and $L$.}
\label{fig:diffCmaxmincautics}
\end{figure}

With the prescription (\ref{deltav}) and the characteristics of the flows presented in \cite{Duffy2008}, the incoming axion momentum distribution is shown on the left of figure~\ref{fig:causticmap}. With this distribution, the same calculations as in the previous section have been performed. As an illustration, on the right of figure~\ref{fig:causticmap} the form factor (\ref{fk}) for this model, $|\mathcal{F}_{\mathrm{IM}}|^2$, is plotted for any orientation of the cavity, for an example case of $m_a=7\times 10^{-5}$ eV and $L=15$ m. It shows an almost full variation between the orientations perpendicular and parallel to the Big Flow. Figure~\ref{fig:maxCcaustics} shows the maximum value $|\mathcal{F}_{\mathrm{IM}}|^2$ for a range of values of $m_a$ and $L$, while figure~\ref{fig:diffCmaxmincautics} shows the difference between maximum and minimum of $|\mathcal{F}_{\mathrm{IM}}|^2$ for that same range of values of $m_a$ and $L$. Compared with the equivalent figures of the isothermal sphere halo model (figures~\ref{fig:maxC} and \ref{fig:diffCmaxmin}) one can see that the coherence is preserved for longer cavity lengths, provided the cavity is properly oriented (usually perpendicular to the main stream direction). Also the modulation effect between the orientation perpendicular and parallel is larger in this case, and also happens at longer cavity lengths.
To conclude, although the quantitative details depend on the specificities of the model, and in particular on the velocity dispersion of the stream(s), the presence of low dispersion streams in the axion dark matter distribution will enhance the effect studied here by: 1) enhancing the difference of the expected signal for different orientations of the cavity lengths (directional modulation); and 2) enhancing the expected signal even for cavities longer than the axion de Broglie wavelength, given proper cavity orientation. In general, in case of a positive detection a detailed study of the signal dependence with cavity orientation will give information on type, number and dispersion of dark matter streams, much like the conceptually similar WIMP directional experiments.

\section{Sensitivity}
\label{sec:sensitivity}

We now address the issue whether realistic values for axion model parameters with interest as axion dark matter can give detectable signals in cavity geometries which follow the prescriptions developed in previous section, aimed at observing a possible directional dependence.
In order to answer this, we have performed an approximate sensitivity calculation of the axion-photon coupling $g_{a\gamma}$ for a range of axion mass $m_a$ from $10^{-5}$ eV to $1.5\times10^{-4}$ eV, for some a priori reasonable experimental parameters and cavity geometries consistent with the requirements exposed above.

The computation is done in the usual way for haloscopes~\cite{Asztalos:2001tf} but using the signal strength according to~(\ref{pout2}) for the cavity orientation giving the maximum signal and with cavity geometries fixed by the needed resonance to $m_a$. We assume all local dark matter density is composed by axions and is equal to 0.3 GeV cm$^{-3}$. We focus on the velocity distribution of the isothermal sphere model of section~\ref{sec:halo}, knowing that the eventual presence of streams will largely improve our conclusions.

\begin{table}[t]
  \centering
\begin{center}
\begin{tabular}[b]{ccccc}
\hline
Axion mass (eV) & Length (m) & $|\mathcal{F}_{IS}^{\bf{max}}|^{2}$ & $|\mathcal{F}_{IS}^{\bf{max}}|^{2}$ - $|\mathcal{F}_{IS}^{\bf{min}}|^{2}$ & Volume (l) \\
\hline
 $2.0\times10^{-5}$ & 20 & 0.95 & 0.13 & 18.59 \\
 $4.0\times10^{-5}$ & 20 & 0.85 & 0.28 & 9.30 \\
 $6.0\times10^{-5}$ & 15 & 0.83 & 0.30 & 4.65 \\
 $8.0\times10^{-5}$ & 15 & 0.75 & 0.35 & 3.49 \\
 $1.0\times10^{-4}$ & 10 & 0.80 & 0.32 & 1.86 \\
 $1.2\times10^{-4}$ & 10 & 0.75 & 0.35 & 1.55 \\
 $1.4\times10^{-4}$ & 10 & 0.70 &0.37  & 1.33 \\
\hline
\end{tabular}
\end{center}
  \caption{Values of the length $L$ used to generate the plots in figure \ref{fig:explot1} for several indicative axion masses, as well as the maximum value of the factor $|\mathcal{F}_{IS}|^{2}$, its modulation (maximum minus minimum), and the cavity volume, corresponding to each case.}\label{tab:values}
\end{table}

\begin{figure}[t]
\centering
\includegraphics[width=11cm]{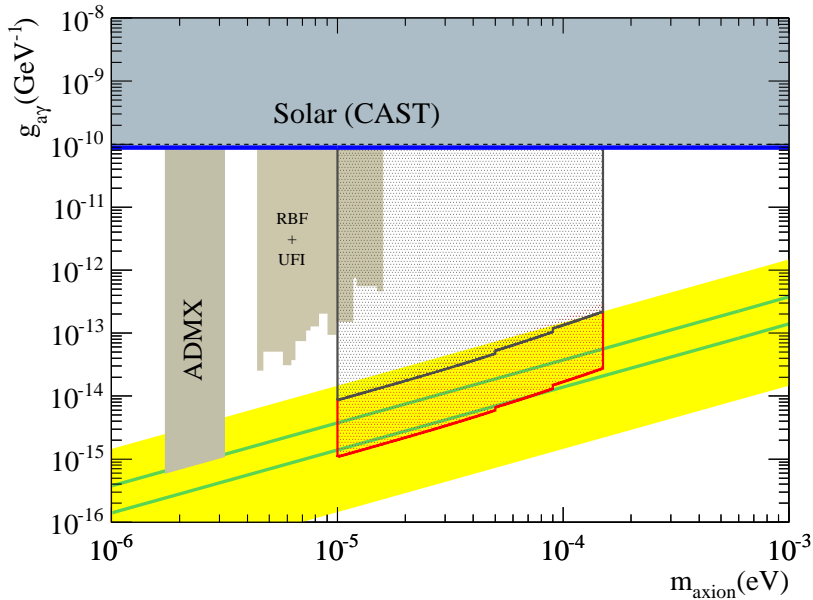}
\caption{Sensitivity regions for the models giving rise to a signal to noise ratio of at least 5 in a cavity geometry fixed by the prescription given in the text to obtain a directional effect (see text for details). The conservative option (black region) is calculated for $Q=1000$, $t_{\bf{step}}=30000 $ s, and rest of input parameters as explained in the text and table~\ref{tab:values}. The more aggressive option (red region) contemplates a better quality factor $Q=20000$, $t_{\bf{step}}=3000$, and a volume 10 times larger that the previous one. Also shown are the areas excluded by current and past axion haloscopes~\cite{Hagmann1990,Wuensch1989,DePanfilis1987,Asztalos:2001tf,Asztalos2010}, as well as the CAST axion helioscope~\cite{Andriamonje:2007ew}. The yellow band indicated the region favored by the QCD axion models, being the green lines two representative benchmark models, the KSVZ~\cite{Kim:1979if,Shifman:1979if} axion (upper one) and the DSFZ~\cite{Dine:1981rt,Zhitnitsky:1980tq} axion (lower one).}
\label{fig:explot1}
\end{figure}

The values of the different experimental parameters entering the sensitivity calculation are fixed after the following considerations. For each axion mass $m_a$ the transversal dimension $d_y$ of the cavity is fixed by the resonance condition. The length $L$ is fixed differently in diverse ranges of $m_a$ so that it lies approximately in the region indicated by the plot~\ref{fig:diffCmaxmin} as having the maximum modulation. The specific values taken for $L$ are listed in table~\ref{tab:values} for several values of the $m_a$. The remaining dimension $d_x$ is fixed at 3 cm for all the calculations. This procedure is an approximation, as in reality the cavities would be built with a system to tune the resonance frequency by means of movable dielectric pieces inside the cavity, like the ones used in ADMX or the ones described in~\cite{Baker:2011na}. In a proper calculation the presence of the dielectric pieces must be taken into account in the calculation of $\tilde{C}$ both due to the geometrical variation of the TE$_{101}$ mode and the presence of dielectric material in the cavity. These issues have been neglected here so we consider this result within a factor $\sim2$ uncertainty. Following similar criteria as in~\cite{Baker:2011na}, the quality factor $Q$ of the cavity has been assumed to be constant and equal to 1000 and the noise figure has been assumed to be 3 K. For every step an integration time of 30000 s is assumed which adds up a total of 2.5 effective years to scan the full mass range indicated in figure~\ref{fig:explot1} from $10^{-5}$ to $1.5\times 10^{-4}$ eV. All these parameters are considered a priori feasible but of course a detailed technical study is needed, something beyond the scope of this paper. First considerations of technical nature are found in~\cite{Baker:2011na}. Finally, we have assumed a magnetic field of 10 T.

The upper line shown in figure \ref{fig:explot1} represents the values of $\gagamma$ giving a signal-to-noise ratio of 5 for the input parameters mentioned. As shown, the expected sensitivity in $\gagamma$ is already at the level of the uppermost values of the realistic QCD axion model band, even for such conservative input parameters. More optimistic values are considered to obtained the lower line, for which a quality factor of 20000 is instead used, as well as 3000 s of integration time (corresponding to 5 years of effective time for the full scan), as well as a $\times$10 larger volume $V$ with respect the conservative case, to be achieved by the use of 10 cavities of the dimensions stated with their power combined coherently. This second line goes deeply into the QCD axion band and in particular widely covers both the KSVZ and DSFZ benchmark models for a good part of the mass range studied. Of course the regions plotted in figure~\ref{fig:explot1} indicate the approximate sensitivity within the time indicated; in the case of a signal is found, then systematic measurements with different cavity orientation should be performed to study the actual signal modulation (like those of figures~\ref{fig:isothermalsphere} or \ref{fig:causticmap}).

In summary, the directional effect discussed above arisen from the orientation-dependent conversion conditions of the relic axions in a long thin cavity, requires cavity length of the order of a few meters, something realistic to implement in current or future dipole magnets. Moreover, the signal intensity in these geometries seems to be sufficiently intense to explore the QCD axion models of interest as dark matter candidates for the mass range of $10^{-5} \sim 10^{-4}$ eV.

\section{Conclusions}

We have studied the expected axion dark matter signal in a resonant cavity of an axion haloscope detector, not satisfying the condition that the axion de Broglie wavelength is sufficiently larger than the cavity dimensions for a fully coherent conversion, i.e. $\lambda_a \gtrsim 2\pi L$. In this case, and for cavity geometries largely non-spherical, the expected signal develops a dependency on the direction of the incoming axion or, equivalently, the orientation of the cavity with respect the distribution of relic axion momenta. This is the first time, to our knowledge, that a directional effect in axion dark matter detection is studied, effect that can be exploited to design directional axion dark matter detectors, conceptually equivalent (although technologically rather different) than the much sought WIMP dark matter directional detectors\footnote{It is interesting to note that WIMP directionality is achieved by measuring the WIMP-induced nuclear recoil direction in a suitable particle detector. It is a completely different mechanism, whose directionality is not dependent on the size of the detector~\cite{Ahlen:2009ev}.}. The directional signal of dark matter, proposed 25 years ago for the first time~\cite{Spergel:1987kx} in the context of WIMP searches, is acknowledged to provide an unmistakable signature of the extraterrestrial origin of a possible positive detection.

We have focused our study to simple long and thin cavity geometries immersed in powerful dipole magnets, like the ones recently proposed to achieve competitive sensitivity in the $10^5-10^4$ eV axion mass range. We found that indeed a O(1) modulation of the signal with respect to the cavity orientation is present if adequate cavity lengths are employed, even when a realistic axion momentum distribution is assumed using the isothermal sphere model for the galactic dark matter distribution. Other halo models will yield a different modulation signals, although we do not expect a difference in the qualitative conclusions here stated for halo models that suppose a smooth departure from the isothermal sphere. For more radically different models, like for example the caustic rings model, this effect is even more identificative and interesting. The presence of low dispersion flows in the dark matter distribution make the effect studied here valid even for high axion masses, pushing the range of validity well above the conventional limitation provided the cavity is oriented perpendicular to the incoming flow direction. In any case, this effect no only provides a clear signature of dark matter but would give extra information on the particular galactic distribution of dark matter, an invaluable information on the galaxy halo formation and evolution.

The case studied involving long cavities in dipole magnets is particularly appealing because it could be realized in the near future given that this type of magnets, developed for accelerator technology, are already in use by the axion community in experiments looking for solar axions, like CAST~\cite{Aune:2011rx}, or axions (or ALPs) produced in the laboratory, like ALPS~\cite{Ehret:2010mh}. These results add motivation to the use of these setups also as haloscopes in search for relic axions. Although the movement of rotation of the Earth could allow a static magnet to scan the sky in search for the modulation studied here, we note that magnets used or foreseen for helioscopes are equipped with movable platforms (to orient the magnet to the Sun) that would also be of additional value in a directional relic axion campaign, providing flexibility to scan the sky more efficiently.

\acknowledgments

We thank F. Caspers and P. Sikivie for interesting and encouraging discussions. We owe to F. Caspers the suggestion of a directional effect in long thin cavities. We thank specially J. Redondo for many suggestions and a careful reading of the manuscript. We acknowledge support from the Spanish Ministry of Science and Innovation (MICINN) under contracts FPA2008-03456 and FPA2011-03456. Part of these grants are funded by the European
Regional Development Fund (ERDF/FEDER). We also acknowledge support from the European Commission under the European Research
Council T-REX Starting Grant under contract number ERC-2009-StG-240054 of the IDEAS program of the 7th EU Framework Program.

\label{sec:conclusions}

\bibliographystyle{JHEP}
\bibliography{../../BibTeX/igorbib}

\end{document}